\documentclass[preprint,12pt,a4paper]{elsarticle}

\usepackage{todonotes}
\usepackage[T1]{fontenc}
\usepackage{amssymb}
\usepackage[dvipsnames]{xcolor}
\usepackage{graphicx}
\usepackage{tikz}
\usetikzlibrary{arrows.meta}
\usepackage{booktabs}
\usepackage{rotating}
\usepackage{hyperref}
\usepackage{tabularx}
\newcolumntype{Y}{>{\raggedright\arraybackslash}X}

\journal{SoftwareX}

\begin{document}
\renewcommand{\labelenumii}{\arabic{enumi}.\arabic{enumii}}

\begin{frontmatter}

\title{AMoRE: An agentic framework for end-to-end metamorphic testing of web applications}

\author[uma]{Andrei Coman\corref{cor1}}
\ead{andreialexcoman@uma.es}
\cortext[cor1]{Corresponding author.}

\author[umu]{Jos\'e Luis Fern\'andez Alem\'an}
\ead{aleman@um.es}

\author[umu]{Bego\~na Moros Valle}
\ead{bmoros@um.es}

\address[uma]{ITIS Software, University of M\'alaga, Spain}
\address[umu]{University of Murcia, Spain}

\begin{abstract}
Web applications are widely used in everyday activities such as e-commerce, banking, transportation, and healthcare, therefore ensuring their reliability is essential. End-to-end testing allows developers to validate complete workflows from the user's perspective, but designing and maintaining these tests remains costly. While interaction flows can be clearly defined, determining their expected outcomes is challenging when interfaces depend on dynamic data, session state, or external services. This challenge, known as the oracle problem, can be addressed through metamorphic testing, which verifies relations between related executions instead of requiring explicit expected results. However, metamorphic relations are typically defined manually and tailored to individual pages, limiting their applicability to generic web interfaces and their scalability in the context of increasingly prevalent AI-generated code. To address these limitations, we present \emph{AMoRE} (Agentic Metamorphic Relation Explorer), an agentic tool for automatically discovering and generating metamorphic tests for websites supplied at test time. AMoRE follows a hybrid architecture that combines vision-language model-based UI exploration with deterministic test compilation, execution, and evaluation. A vision-language model explores the interface using a Set-of-Mark inventory and proposes structured metamorphic relations. These relations are then compiled into Playwright tests and evaluated through paired executions by a deterministic engine, without requiring further model calls. In a preliminary evaluation across five commercial websites, AMoRE generated and compiled 63 metamorphic relations from 100 exploration attempts, with 48 tests executing successfully.
\end{abstract}


\end{frontmatter}


\section*{Metadata}
\begin{table}[!h]
\begin{tabular}{|l|p{5.8cm}|p{5.8cm}|}
\hline
\textbf{Nr.} & \textbf{Code metadata description} & \textbf{Metadata} \\
\hline
C1 & Current code version & v1.0.0 \\
\hline
C2 & Permanent GitHub link to code/repository used for this code version & \url{https://github.com/a-coman/AMoRE} \\
\hline
C3 & Legal Code License & MIT License \\
\hline
C4 & Code versioning system used & git \\
\hline
C5 & Software code languages, tools, and services used & TypeScript on Node.js 22; Next.js and React; NestJS; Playwright; LangGraph; Prisma and PostgreSQL; RabbitMQ; Garage (local S3 compatible API); Docker Compose; OpenRouter API \\
\hline
C6 & Compilation requirements, operating environments \& dependencies & Node.js 22.13 or newer, pnpm 11.5 or newer, Docker Compose, and an OpenRouter API key for LLM-assisted exploration. Tested on Linux and macOS hosts; workers run in Docker containers \\
\hline
C7 & If available Link to developer documentation/manual & \url{https://github.com/a-coman/AMoRE/blob/main/README.md} \\
\hline
C8 & Support email for questions & andreialexcoman@uma.es \\
\hline
\end{tabular}
\caption{Code metadata}
\label{tab:metadata}
\end{table}

\section{Motivation and significance}
\label{sec:motivation}

End-to-end (E2E) tests are the layer of the testing pyramid that checks web applications the way users experience them, by driving a real browser and observing the resulting interface. However, web interfaces are assembled at runtime from asynchronous requests, client-side routing, and third-party components, so the same page can differ between two visits because of locale, session state, or catalogue changes. This variability makes E2E tests costly to write and maintain, particularly when deciding whether an observed outcome is correct. That difficulty is the oracle problem~\cite{Weyuker1982,Barr2015Oracle}. Assertions tied to exact page content or layout can break when a label or presentation changes without any functional fault, while a weak assertion such as ``the page loaded'' can miss real defects~\cite{Selay2025WebTesting,Kaynak2025LLMShot}.

Metamorphic testing (MT) reframes the problem~\cite{Chen1998,Chen2018Survey,Segura2016MTSurvey}. Instead of asking whether one output is correct, it checks a metamorphic relation (MR) between related executions. An MR defines a source execution, a transformation that yields a follow-up execution, and a property that must hold between their observations. For example, under equivalent session and data conditions, adding a restrictive filter to a search should not increase the result count, repeating a search should preserve its observable results, and applying two independent filters in either order should lead to the same set. In MT no exact expected output needs to be specified in advance, only the relation. MT has found faults in search engines, RESTful APIs, and web security mechanisms~\cite{Zhou2016SearchEngines,Segura2018REST,Chaleshtari2023Metamorphic}. However, applying these relations to web interfaces requires page-specific interaction paths and observables. Whoever writes an MR for the web must find a navigable path, locate the controls, decide which parts of the page state to observe, and express all of it as actions that can be repeated. These details depend on the site and drift with its DOM, its asynchronous loading, and its promotional overlays. MT therefore does not remove the effort needed to construct executable MRs for a specific URL supplied at test time. Nowadays, agents based on large language models (LLMs) already guide GUI exploration and test generation~\cite{Liu2024GPTDroid,Wang2024Software}, while vision-language models can use screenshots to interpret the interface and detect functional problems~\cite{Liu2025Seeing}. This raises the question of whether such an agent can discover executable MRs on a web interface without site-specific configuration, and whether those MRs stay useful once compiled, reviewed, and replayed.

Table~\ref{tab:related} compares AMoRE with the closest prior work identified in a systematic mapping study we conducted, covering 30 studies on MT, LLM-assisted testing, and web or GUI automation. We searched the ACM Digital Library, IEEE Xplore, and arXiv and added 12 seed references, which gave 389 records and 276 after deduplication. An LLM-assisted abstract screening followed by a manual triage retained the 30 studies that were read in full. The database exports, the screening prompt, and the screening toolkit are available in the repository~\cite{AMoRERepo}. The table compares the approaches along four dimensions, namely the system under test, the source of the MRs, the artifact the approach produces, and what decides the final verdict. Existing approaches cover complementary parts of the problem. Segura et al.\ define reusable MR patterns for query-based systems~\cite{Segura2019MRP}, and MST-wi compiles 76 security MRs written in a domain-specific language into Java tests~\cite{Chaleshtari2023Metamorphic}. LLM-based approaches derive MRs from requirements documents, API specifications, or manuals~\cite{Zhang2023ChatGPTMR,Shin2024Towards,Tsigkanos2023Large,Hazott2025LLM}. Other approaches, such as MR-Scout, mine and synthesize relations from existing test suites~\cite{Xu2024MR}. Multi-agent pipelines such as ARMeta compile generated MRs into executable tests, but for REST APIs rather than browser interfaces~\cite{Khan2026Multi}.

GUI agents address the exploration side. GPTDroid guides mobile GUI interaction through textual descriptions of the interface and evaluates its ability to detect crashes~\cite{Liu2024GPTDroid}. VisionDroid uses multimodal agents to explore mobile applications and identify non-crash functional bugs through model-based reasoning~\cite{Liu2025Seeing}. Neither approach produces metamorphic relations. Assure applies MT to AI-powered browser extensions through predefined relations and synthetic page families~\cite{Gao2025Assure}. Finally, 22 of the 30 studies in the mapped corpus use MT to evaluate LLMs or LLM-based applications rather than conventional software. Where an LLM acts as the final oracle, its output variability also complicates its use for regression testing~\cite{Dobslaw2026Challenges}. No study in the corpus combines MR discovery through exploration of a web interface supplied at runtime, compilation into replayable browser tests, and a verdict independent of the LLM.

\begin{table*}[t]
\centering
\caption{AMoRE compared with the closest prior work.}
\label{tab:related}

\footnotesize
\renewcommand{\arraystretch}{1.18}
\setlength{\tabcolsep}{4pt}

\begin{tabularx}{\textwidth}{@{}YYYYY@{}}
\toprule
\textbf{Approach} &
\textbf{System under test} &
\textbf{MR source} &
\textbf{Produced artifact} &
\textbf{Verdict} \\
\midrule

MR patterns~\cite{Segura2019MRP} &
Query-based systems &
Manual pattern catalogue &
Reusable MR schemas &
Not prescribed \\

\addlinespace[3pt]

MST-wi~\cite{Chaleshtari2023Metamorphic} &
Web security &
Manual DSL catalogue &
Java tests, crawled inputs &
Deterministic \\

\addlinespace[3pt]

Shin et al.~\cite{Shin2024Towards} &
Web security &
LLM from requirements text &
DSL and Java &
Deterministic \\

\addlinespace[3pt]

ARMeta~\cite{Khan2026Multi} &
REST APIs &
Multi-agent from OpenAPI &
Behave tests &
Deterministic \\

\addlinespace[3pt]

GPTDroid~\cite{Liu2024GPTDroid} &
Mobile GUIs &
None (no MT) &
Interaction commands and bug reports &
Crash detection \\

\addlinespace[3pt]

VisionDroid~\cite{Liu2025Seeing} &
Mobile GUIs &
None (no MT) &
Functional bug reports &
Model judgement \\

\addlinespace[3pt]

Assure~\cite{Gao2025Assure} &
AI browser extensions &
Predefined relations, synthetic pages &
Scripts &
DOM and semantic checks \\

\midrule

\textbf{AMoRE} &
\textbf{Web UIs supplied as URLs} &
\textbf{LLM exploration over a runtime inventory} &
\textbf{Playwright playbooks} &
\textbf{Deterministic} \\

\bottomrule
\end{tabularx}
\end{table*}

AMoRE~\cite{AMoRERepo} addresses this gap with a framework that accepts a user-supplied URL, explores the rendered page with a bounded LLM agent, drafts candidate MRs from four transformation families (idempotence, subset, permutation, and inverse)~\cite{Segura2016MTSurvey,Segura2019MRP}, compiles them into Playwright tests, and evaluates paired source and follow-up runs with a verdict computed by deterministic code. The LLM is used during exploration and drafting, while compilation, replay, and final evaluation never call it. Additionally, a web interface lets a tester inspect each proposed relation together with screenshots and browser traces, edit its compiled playbook, and approve or reject the draft. Approved relations can then be replayed as Playwright tests.

\section{Software description}
\label{sec:description}

The design had to meet seven requirements drawn from the gap above, namely (R1) accept any URL without site-specific adapters; (R2) discover relations from the rendered page; (R3) keep the LLM output structured and compilable; (R4) add a human in the loop to approve a draft before it becomes a test; (R5) replay approved relations without LLM inference; (R6) keep every screenshot, trace, observation, and verdict traceable; and (R7) scale browser automation and LLM inference independently so that several sessions can run at once.

\subsection{Software architecture}
\label{sec:architecture}

\begin{figure}[!h]
  \centering
  \includegraphics[width=\linewidth]{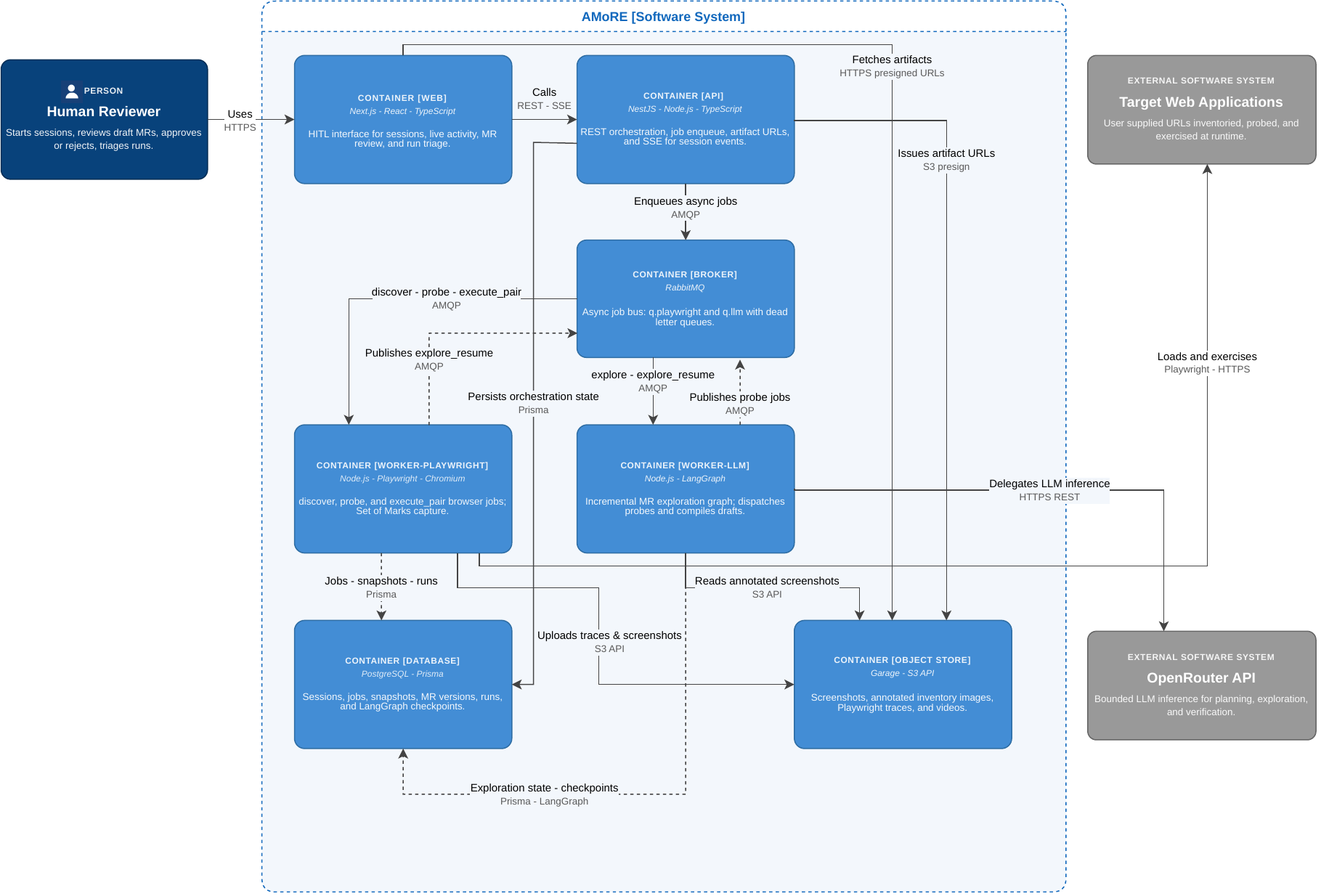}
  \caption{C4 container view of AMoRE.}
  \label{fig:architecture}
\end{figure}

AMoRE is a TypeScript monorepo organised into deployable applications and shared packages. Figure~\ref{fig:architecture} shows the architecture container view following the C4 model proposed by Simon Brown~\cite{brown2026c4model}. A Next.js web application is the human in the loop (HITL) interface. A NestJS API owns session lifecycle, permissions, and job orchestration. Two worker pools do the heavy work. The Playwright worker~\cite{PlaywrightDocs} loads pages, builds page inventories, executes exploration probes, and runs approved tests. The LLM worker runs a LangGraph state machine~\cite{LangGraphDocs} that plans and verifies exploration and calls a vision language model through OpenRouter (an OpenAI compatible API). The API and the workers communicate through RabbitMQ over two durable queues, one for browser jobs (discover, probe, execute pair) and one for LLM jobs (plan, explore, verify, observe). PostgreSQL, accessed through Prisma, stores sessions, jobs, page snapshots, MR versions, runs, observations, verdicts, and an audit entry for every model call. Garage (S3 compatible API) stores screenshots and Playwright traces, which the web interface fetches through presigned URLs. Everything runs under Docker.

The code is divided into four main apps (api, web, worker-llm, and worker-playwright) and seven shared packages. The core package holds the MR schemas, the relation engine, and the compiler that turns exploration slots into Playwright playbooks. The API and both workers consume it, so exploration, approval, and replay apply the same rules. An inventory package implements Set-of-Mark (SoM) capture~\cite{yang2023som} over Playwright, and a contracts package validates every RabbitMQ message.

The architecture follows one principle: the LLM proposes and the rest of the system decides. The model fills structured slots whose element references must resolve against a SoM inventory captured at runtime, so it never emits free selectors or arbitrary code. The compiler turns those slots into executable code, the Playwright worker alone runs it, and the relation engine alone produces the verdict. Separating browser automation from model inference also lets each worker pool scale on its own and lets one session explore all four relation families in parallel.

\subsection{Software functionalities}
\label{sec:functionalities}


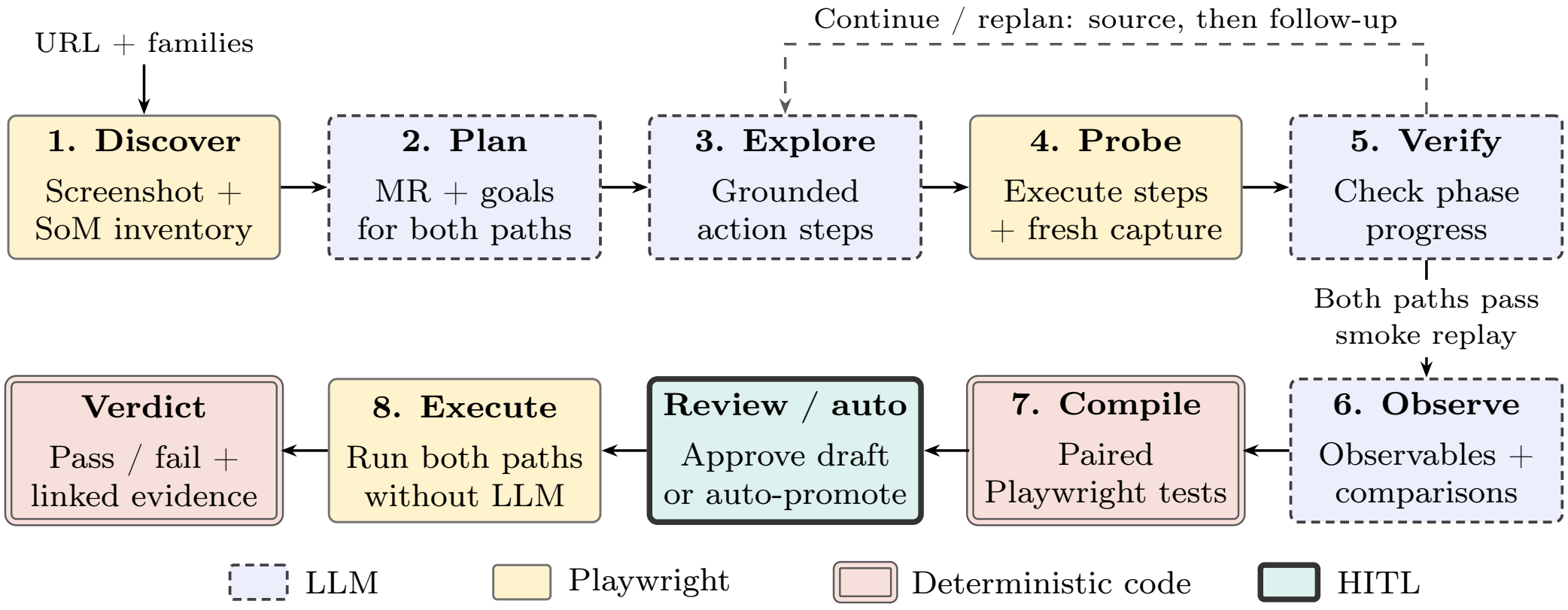
\begin{figure}[htbp]
  \centering
\begingroup
\definecolor{workflowLLM}{HTML}{ECEFFB}
\definecolor{workflowBrowser}{HTML}{FFF3CD}
\definecolor{workflowCode}{HTML}{F7E0DB}
\definecolor{workflowReview}{HTML}{DDF2EF}
\resizebox{\linewidth}{!}{%
\begin{tikzpicture}[
  font=\fontsize{8}{9.5}\selectfont,
  stage/.style={
    draw=black!55,
    rounded corners=2pt,
    line width=0.55pt,
    text width=2.16cm,
    minimum height=1.25cm,
    inner sep=3pt,
    align=center
  },
  llmborder/.style={draw=black!75,line width=0.7pt,dash pattern=on 2.2pt off 1.4pt},
  browserborder/.style={draw=black!55,line width=0.55pt},
  codeborder/.style={draw=black!65,line width=0.45pt,double=workflowCode,double distance=1.1pt},
  reviewborder/.style={draw=black!80,line width=1.4pt},
  llm/.style={stage,fill=workflowLLM,llmborder},
  browser/.style={stage,fill=workflowBrowser,browserborder},
  code/.style={stage,fill=workflowCode,codeborder},
  review/.style={stage,fill=workflowReview,reviewborder},
  flow/.style={-{Stealth[length=4pt,width=3pt]},line width=0.65pt},
  feedback/.style={flow,dashed,draw=black!65},
  note/.style={font=\fontsize{7.5}{9}\selectfont,align=center,fill=white,inner sep=2pt},
  key/.style={minimum width=0.5cm,minimum height=0.3cm,inner sep=0pt,rounded corners=1pt}
]
  \node[browser] (discover) at (0,0)
    {\textbf{1. Discover}\\[3pt]Screenshot +\\SoM inventory};
  \node[llm] (plan) at (2.8,0)
    {\textbf{2. Plan}\\[3pt]MR + goals\\for both paths};
  \node[llm] (explore) at (5.6,0)
    {\textbf{3. Explore}\\[3pt]Grounded\\action steps};
  \node[browser] (probe) at (8.4,0)
    {\textbf{4. Probe}\\[3pt]Execute steps\\+ fresh capture};
  \node[llm] (verify) at (11.2,0)
    {\textbf{5. Verify}\\[3pt]Check phase\\progress};

  \node[llm] (observe) at (11.2,-2.3)
    {\textbf{6. Observe}\\[3pt]Observables +\\comparisons};
  \node[code] (compile) at (8.4,-2.3)
    {\textbf{7. Compile}\\[3pt]Paired\\Playwright tests};
  \node[review] (review) at (5.6,-2.3)
    {\textbf{Review / auto}\\[3pt]Approve draft\\or auto-promote};
  \node[browser] (execute) at (2.8,-2.3)
    {\textbf{8. Execute}\\[3pt]Run both paths\\without LLM};
  \node[code] (verdict) at (0,-2.3)
    {\textbf{Verdict}\\[3pt]Pass / fail +\\linked evidence};

  \node[note] (input) at (0,1.25) {URL + families};
  \draw[flow] (input.south) -- (discover.north);
  \draw[flow] (discover) -- (plan);
  \draw[flow] (plan) -- (explore);
  \draw[flow] (explore) -- (probe);
  \draw[flow] (probe) -- (verify);

  \draw[feedback] (verify.north) -- (11.2,1.25)
    -- node[above,note] {Continue / replan: source, then follow-up}
    (5.6,1.25) -- (explore.north);
  \draw[flow] (verify.south) -- node[note,text width=2.25cm]
    {Both paths pass\\smoke replay} (observe.north);

  \draw[flow] (observe) -- (compile);
  \draw[flow] (compile) -- (review);
  \draw[flow] (review) -- (execute);
  \draw[flow] (execute) -- (verdict);

  \node[key,fill=workflowLLM,llmborder] (kLLM) at (1.0,-3.45) {};
  \node[anchor=west] at (kLLM.east) {LLM};
  \node[key,fill=workflowBrowser,browserborder] (kBrowser) at (3.3,-3.45) {};
  \node[anchor=west] at (kBrowser.east) {Playwright};
  \node[key,fill=workflowCode,codeborder] (kCode) at (6.3,-3.45) {};
  \node[anchor=west] at (kCode.east) {Deterministic code};
  \node[key,fill=workflowReview,reviewborder] (kReview) at (10.0,-3.45) {};
  \node[anchor=west] at (kReview.east) {HITL};
\end{tikzpicture}%
}
\endgroup
  \caption{AMoRE workflow from URL to verdict.}
  \label{fig:workflow}
\end{figure}

Exploration is organised around four transformation families with bounded comparison operators. Based on~\cite{Segura2019MRP}, we selected (i) idempotence, which repeats an action and expects the observables to stay equal; (ii) subset, which adds a restriction and expects counts not to grow; (iii) permutation, which applies two independent actions in both orders and expects equal sets; and (iv) inverse, which applies an action, undoes it, and expects the earlier state back. The comparison operators are \texttt{equal}, \texttt{not\_equal}, \texttt{cardinality\_lte}, and \texttt{set\_equal}. Family profiles constrain which comparisons the model may choose, which keeps discovery goals aligned with evaluation semantics and avoids open-ended relation invention.

Figure~\ref{fig:workflow} summarises a session from URL to verdict. First, discovery (step 1) loads the URL, waits for the page to settle, and captures a page inventory. Following SoM prompting~\cite{yang2023som}, every interactable candidate receives a short identifier (\texttt{E1}, \texttt{E2}, and so on) that is painted on the screenshot and listed in a textual inventory with its role, accessible name, and visible text. 
Playwright accessibility snapshots are merged with DOM candidates ranked by visibility and stability signals such as \texttt{data-testid} and \texttt{aria-label}, direct controls are ranked ahead of links, and the list is capped so that prompts stay bounded. 

In the plan step (step 2), the LLM proposes a candidate MR for each family and splits it into a source goal and a follow-up goal. The LangGraph graph then runs an explore, probe, and verify loop (steps 3 to 5). The explore step plans one to three steps that reference inventory identifiers, with at most one state-changing action (\texttt{click}, \texttt{fill}, \texttt{press}, \texttt{scroll}, \texttt{select}, \texttt{goto}) per batch. In the probe step the Playwright worker replays the validated prefix plus the new batch and returns a fresh screenshot and inventory. In the verify step the graph checks whether the goal was reached and commits the step or backtracks. When a phase reaches its goal, a smoke replay re-executes the whole path from the start page as a stability gate. If a step fails, the worker first tries alternative locators; if that also fails, the LLM receives the Playwright error and the new screenshot and must propose a different approach. This is how exploration clears cookie banners, promotional modals, and login walls. It does not try to pass CAPTCHA or anti-bot gates, and it abandons an attempt after a budget of 30 iterations is exhausted or when it declares the goal unreachable.

Once both paths pass their smoke replay, the observe step (step 6) uses a separate observation inventory that records stable anchors such as result counts, headings, and URL parts. The model binds each observable to an element of the final source snapshot and picks a comparison allowed by the family. In the compile step (step 7), the compiler validates that every step and binding resolves against the inventory, maps them to Playwright locators, and renders one specification with a shared observation extractor and separate source and follow-up scenarios, together with a JSON Schema for the observation payload.

In review mode the compiled MR and test waits for a person to approve it. The session hub UI shows per-family progress and a live activity timeline. The MR page shows the exploration timeline with annotated screenshots and trace links, the table of observables and comparisons, and an editor for the compiled playbook. The reviewer can patch the playbook, approve, or reject. Auto mode promotes compiled drafts immediately, which is what the evaluation used to measure feasibility. Long explorations can be paused and resumed cooperatively without losing state.

Approved relations run as paired executions (step 8). The Playwright worker runs the source and follow-up scenarios with tracing enabled and no retries, extracts the observation payloads, and hands them to the relation engine, which compares observable by observable and produces the verdict, reporting pass only if every comparison holds. We call this all-or-nothing verdict strict, since a single failing comparison fails the whole pair. A failure creates a violation linked to the run, its screenshots, and its traces. Nothing in this stage calls the model, so repeated runs measure the stability of the test and of the application rather than the variance of an LLM judge.

\section{Illustrative examples}
\label{sec:examples}

\subsection{A running example on GitHub search}

\begin{figure}[htbp]
  \centering
  \includegraphics[width=\linewidth]{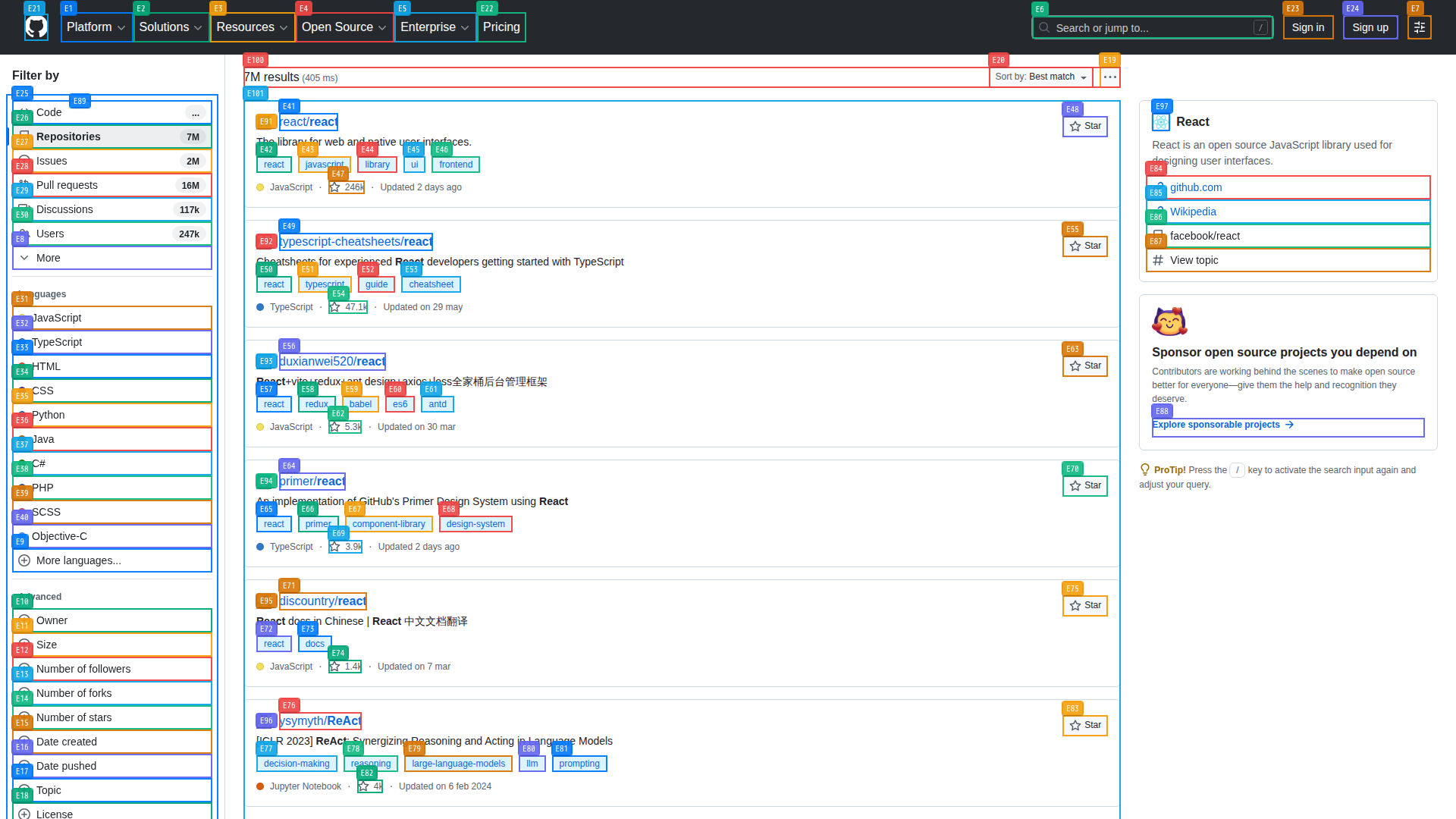}
  \caption{SoM inventory of a GitHub search results page. Each labelled element is listed in a textual inventory that the LLM must use when proposing steps.}
  \label{fig:inventory}
\end{figure}

Consider a tester who wants regression checks for GitHub's global search and knows nothing about its DOM. They open the AMoRE dashboard, enter \url{https://github.com}, select the subset family, and choose review mode. The Playwright worker loads the home page and captures the inventory shown in Figure~\ref{fig:inventory}, where the header search box, the result-type filters on the left, and the sort control each receive an identifier.

\begin{figure}[htbp]
  \centering
  \begin{minipage}[t]{0.49\linewidth}
    \centering
    \includegraphics[width=\linewidth]{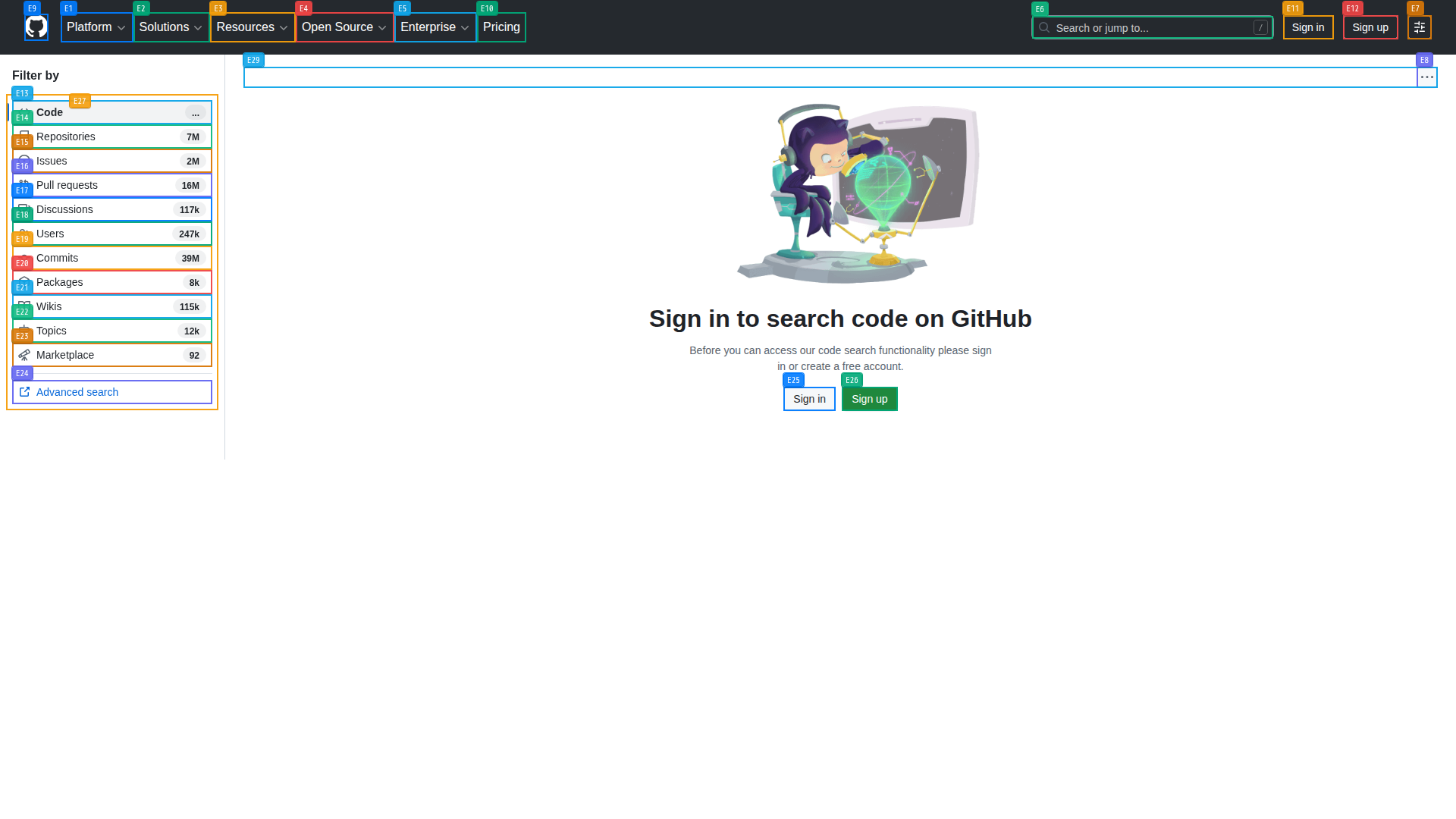}
    \par\vspace{0.3em}(a) Login wall after selecting the Code filter.
  \end{minipage}\hfill
  \begin{minipage}[t]{0.49\linewidth}
    \centering
    \includegraphics[width=\linewidth]{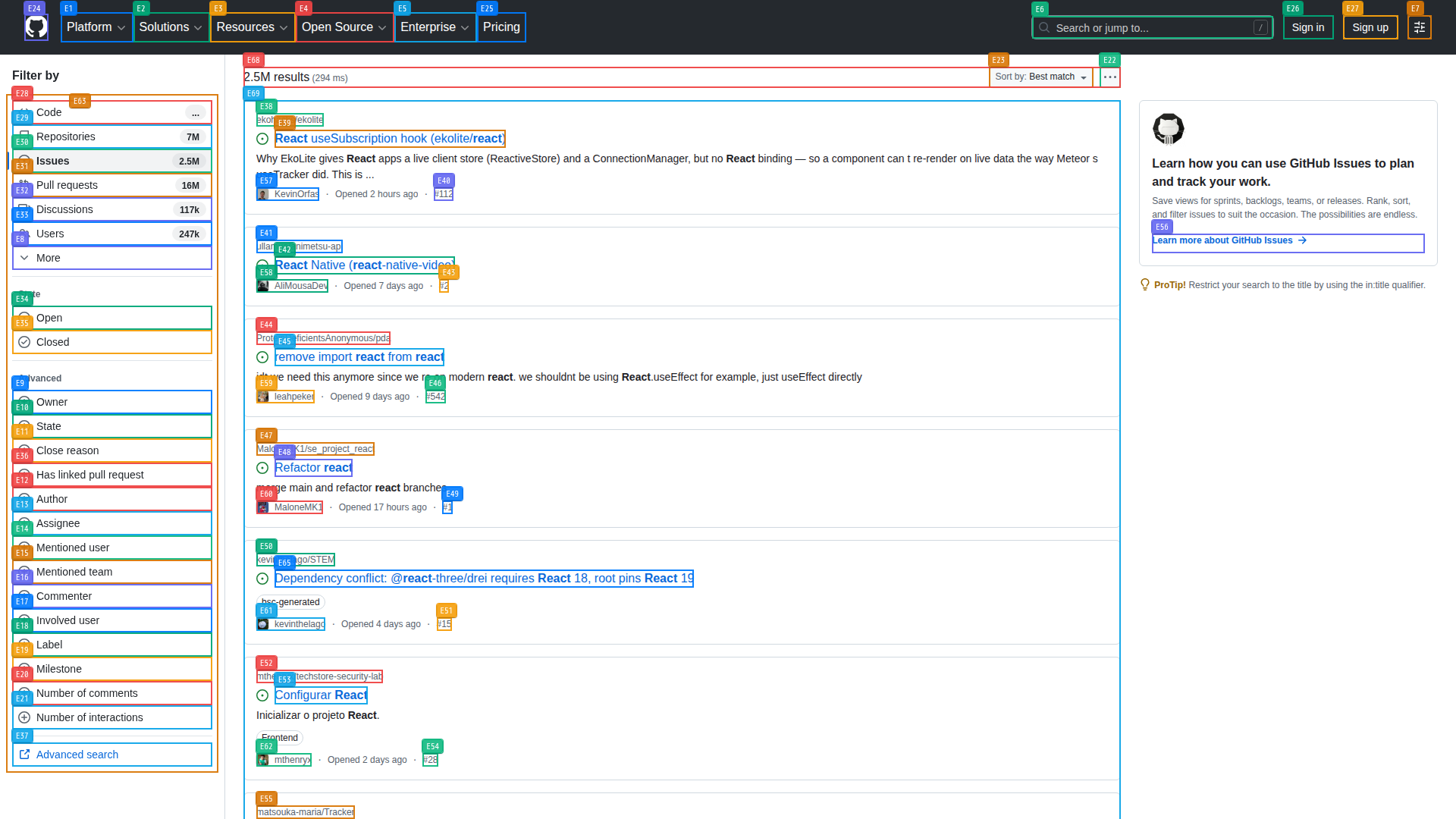}
    \par\vspace{0.3em}(b) Public Issues view after self-healing.
  \end{minipage}
  \caption{Recovery from a login wall during the follow-up phase of a subset relation on GitHub.}
  \label{fig:healing}
\end{figure}

The LLM proposes the relation ``adding a filter to a search must not increase the number of results'' and splits it into a source goal, reach the results page for one query, and a follow-up goal, rebuild that page and apply one extra restriction. Exploring the source path takes two validated steps, click the search control and fill, and lands on a repositories view reporting about seven million results. The follow-up path repeats those steps and then has to narrow the set. The first plan selected the Code filter. The click succeeded, but the verify step found a ``Sign in to search code'' page and no result count (Figure~\ref{fig:healing}a), so the goal was not met. The next plan chose the publicly reachable Issues filter, which produced 2.5 million results with visible listings (Figure~\ref{fig:healing}b). 

The model then selects observables. For this family they typically include the URL pathname and the base query, which must stay equal and the reported result count, which must not grow. The compiler produces a Playwright specification with two scenarios and a schema for these values. The reviewer now sees the relation, the timeline with the login wall and the recovery, and the playbook, and this is where review earns its place. In the evaluation, several GitHub subset drafts compared the complete \texttt{q} parameter with equality even though GitHub appends the filter qualifier to that same parameter, and some parsed the label ``7M'' as the integer 7. A reviewer spots this in the observation table, drops or redirects the comparison in the editor, and approves. From then on the pair replays in about 24 seconds, the study median, with no model call, and each run leaves screenshots, traces, and observation values behind for triage.

\subsection{Evaluation on five commercial websites}

We ran AMoRE in auto mode on Amazon, Booking, Airbnb, MediaMarkt, and GitHub with all four families and five generations per site, which gives 100 exploration attempts. All attempts used MiniMax-M3~\cite{MiniMaxM3}, an open-weight mixture-of-experts vision-language model with 428B total and 23B active parameters, accessed through OpenRouter with a Spanish browser locale over three days in July 2026. In preliminary experiments, some sites returned CAPTCHA or access-denied pages to the headless browser (Decathlon, Fnac, El Corte Ingl\'es, InfoJobs), so they were excluded beforehand. To support reproducibility, we follow the guidelines of Baltes et al.\ for empirical studies involving LLMs~\cite{Baltes2026Guidelines}. Table~\ref{tab:llmconfig} reports the model identifier and snapshot, sampling parameters, memory, and run dates, and the prompts, the exploration graph, and the validation scripts with their outputs are published in the repository~\cite{AMoRERepo}.

\begin{table}[htbp]
  \centering
  \caption{LLM configuration.}
  \label{tab:llmconfig}
  \footnotesize
  \begin{tabularx}{\linewidth}{@{}lY@{}}
    \toprule
    Model & MiniMax-M3 (428B total, 23B active parameters)~\cite{MiniMaxM3} \\
    Identifier & \texttt{minimax/minimax-m3} \\
    Snapshot & \texttt{minimax-m3-20260531} \\
    Access & OpenRouter, default provider routing \\
    Price & USD \texttt{0.30} input and \texttt{1.20} output per million tokens \\
    Temperature$^\dagger$ & OpenRouter default (\texttt{1.0}) \\
    TopP$^\dagger$ & OpenRouter default (\texttt{0.95}) \\
    Reasoning$^\dagger$ & Model default (\texttt{adaptive thinking}) \\
    Dates & 9--11 July 2026, commit \texttt{f3ec9d5} \\
    Memory & OpenRouter default (max \texttt{1048576} tokens)$^{*}$. No chat history. Each call is rebuilt from the LangGraph state, checkpointed in PostgreSQL, with the MR goal, the last 10 batches of the phase, and the current SoM inventory and screenshot \\
    \bottomrule
  \end{tabularx}
  \par\smallskip
  {\scriptsize\raggedright $^\dagger$ We did not set these parameters in the API requests, so the listed values are the defaults that applied.\par
  $^{*}$ Only planning calls set a limit of 4,096 output tokens.\par}
\end{table}

Exploration compiled 63 of the 100 attempts into Playwright tests, and 48 of those completed their first paired execution. Figure~\ref{fig:heatmap} shows how much this depends on site and family. MediaMarkt compiled 80\% of attempts and Booking 15\%; idempotence compiled 80\% and permutation 36\%. Repeating an action or adding one filter needs one stable control, whereas a permutation needs two independent controls that stay available in both orders, which dynamic e-commerce filters rarely offer. Across all sessions only 5.1\% of the 1,833 browser probes failed, and 2.5\% within attempts that eventually compiled, so replanning absorbed most local obstacles. Compiled paths were short, with a median of two validated source steps and four follow-up steps.

\begin{figure}[htbp]
  \centering
  \begin{tikzpicture}[x=1.75cm,y=0.72cm]
  \node[font=\scriptsize,align=left] at (0,0.65) {\textbf{Domain}};
  \node[font=\scriptsize,align=center] at (1,0.65) {\textbf{Idempotence}};
  \node[font=\scriptsize,align=center] at (2,0.65) {\textbf{Subset}};
  \node[font=\scriptsize,align=center] at (3,0.65) {\textbf{Permutation}};
  \node[font=\scriptsize,align=center] at (4,0.65) {\textbf{Inverse}};

    \foreach \name/\row in {
      Amazon/0,
      Airbnb/-1,
      Booking/-2,
      GitHub/-3,
      MediaMarkt/-4
    }{
      \node[
        anchor=west,
        font=\scriptsize
      ] at (-0.6,\row) {\name};
    }

  \fill[RoyalBlue!85] (0.55,-0.31) rectangle (1.45,0.31);
  \node[text=white,font=\scriptsize] at (1,0) {100\%};
  \fill[RoyalBlue!52] (1.55,-0.31) rectangle (2.45,0.31);
  \node[font=\scriptsize] at (2,0) {60\%};
  \fill[RoyalBlue!36] (2.55,-0.31) rectangle (3.45,0.31);
  \node[font=\scriptsize] at (3,0) {40\%};
  \fill[RoyalBlue!85] (3.55,-0.31) rectangle (4.45,0.31);
  \node[text=white,font=\scriptsize] at (4,0) {100\%};

  \fill[RoyalBlue!70] (0.55,-1.31) rectangle (1.45,-0.69);
  \node[text=white,font=\scriptsize] at (1,-1) {80\%};
  \fill[RoyalBlue!70] (1.55,-1.31) rectangle (2.45,-0.69);
  \node[text=white,font=\scriptsize] at (2,-1) {80\%};
  \fill[RoyalBlue!36] (2.55,-1.31) rectangle (3.45,-0.69);
  \node[font=\scriptsize] at (3,-1) {40\%};
  \fill[RoyalBlue!70] (3.55,-1.31) rectangle (4.45,-0.69);
  \node[text=white,font=\scriptsize] at (4,-1) {80\%};

  \fill[RoyalBlue!36] (0.55,-2.31) rectangle (1.45,-1.69);
  \node[font=\scriptsize] at (1,-2) {40\%};
  \fill[black!6] (1.55,-2.31) rectangle (2.45,-1.69);
  \node[font=\scriptsize] at (2,-2) {0\%};
  \fill[black!6] (2.55,-2.31) rectangle (3.45,-1.69);
  \node[font=\scriptsize] at (3,-2) {0\%};
  \fill[RoyalBlue!20] (3.55,-2.31) rectangle (4.45,-1.69);
  \node[font=\scriptsize] at (4,-2) {20\%};

  \fill[RoyalBlue!85] (0.55,-3.31) rectangle (1.45,-2.69);
  \node[text=white,font=\scriptsize] at (1,-3) {100\%};
  \fill[RoyalBlue!85] (1.55,-3.31) rectangle (2.45,-2.69);
  \node[text=white,font=\scriptsize] at (2,-3) {100\%};
  \fill[RoyalBlue!52] (2.55,-3.31) rectangle (3.45,-2.69);
  \node[font=\scriptsize] at (3,-3) {60\%};
  \fill[RoyalBlue!36] (3.55,-3.31) rectangle (4.45,-2.69);
  \node[font=\scriptsize] at (4,-3) {40\%};

  \fill[RoyalBlue!70] (0.55,-4.31) rectangle (1.45,-3.69);
  \node[text=white,font=\scriptsize] at (1,-4) {80\%};
  \fill[RoyalBlue!85] (1.55,-4.31) rectangle (2.45,-3.69);
  \node[text=white,font=\scriptsize] at (2,-4) {100\%};
  \fill[RoyalBlue!36] (2.55,-4.31) rectangle (3.45,-3.69);
  \node[font=\scriptsize] at (3,-4) {40\%};
  \fill[RoyalBlue!85] (3.55,-4.31) rectangle (4.45,-3.69);
  \node[text=white,font=\scriptsize] at (4,-4) {100\%};

\end{tikzpicture}
  \caption{Compilation success rate by site and transformation family. Each cell contains five attempts.}
  \label{fig:heatmap}
\end{figure}

A compiled relation took a median of 10.5 minutes of exploration (IQR 7.5 to 14.3) and 19.3 model calls. The 63 compiled relations cost \$3.17 in model usage, about five cents each, and the full batch including the 37 failed attempts cost \$5.58. Paired execution (source and follow-up) took a median of 24.3 seconds. We then selected 35 compiled relations, two per site and family where available, and replayed each twice more. 23 completed all three runs and 16 reproduced identical observations and verdicts every time. In five of the other seven the observations drifted but the verdict stayed the same. In the remaining two the verdict changed because the observations did, as an Amazon permutation alternated between filter states and a Booking inverse path reached a different page in the replays. No verdict changed while the observations stayed identical, which is consistent with a deterministic relation engine. GitHub was stable in seven of eight sets and idempotence in seven of ten, while MediaMarkt catalogue totals and Airbnb recommendation grids drifted between runs.

Of the 48 compiled relations whose first paired execution completed, 22 passed every comparison and 26 failed at least one, although 80.3\% of the 319 individual observable comparisons passed. We reviewed each of the 26 failures manually against screenshots and traces (Table~\ref{tab:triage}). None was a confirmed defect in the target website. Six correctly rejected pairs whose paths were not equivalent, for example because autocomplete changed the query. Nine came from an invalid relation or a comparison in the wrong direction, such as requiring filter chips to remain after removing a filter. Ten came from brittle observables such as a missing sort label or responsive header chrome. The strongest relations combined query or URL state, normalised counts, and non-empty content, and several of them replayed identically three times. Auto mode was the right choice for measuring feasibility, but a reviewer would have rejected or repaired the misaligned relations and weak observables before they ever ran.

\begin{table}[htbp]
  \centering
  \caption{Manual classification of the 26 strict failures in the initial paired executions.}
  \label{tab:triage}
  \begin{tabular}{@{}p{0.34\linewidth}rp{0.5\linewidth}@{}}
    \toprule
    \textbf{Primary cause} & \textbf{Count} & \textbf{Meaning} \\
    \midrule
    Non-equivalent source and follow-up pair & 6 & The verdict correctly rejected a path that did not realise the intended input relation \\
    Invalid relation or comparison logic & 9 & The relation assumed independent actions that were not, or encoded an expected change as equality \\
    Brittle or incidental observable & 10 & A missing label, dynamic list, capitalisation change, or responsive chrome caused the failure \\
    Unstable or ambiguous evidence & 1 & Replay alternated between equivalent and different filter states \\
    \midrule
    \textbf{Total} & \textbf{26} & \\
    \bottomrule
  \end{tabular}
\end{table}

\section{Impact}
\label{sec:impact}

End-to-end web testing entails substantial development, maintenance, and execution costs. In an industrial case, building a Selenium suite of only 11 test cases took 27.5 hours of manual effort including learning and set-up~\cite{Bons2023Scripted}, and E2E tests take much longer to run and consume far more resources than unit tests~\cite{SotoSanchez2022Dataset}. Moreover, defining and maintaining suitable test oracles remains a major obstacle to test automation. These costs motivate approaches such as metamorphic testing, which can derive test verdicts from relations among multiple executions rather than requiring an explicit expected output for each individual test case.

The main effect of AMoRE is on the effort spent at the top of the testing pyramid. Writing an E2E test for a dynamic page means finding a path, choosing locators, and, hardest of all, deciding what the correct outcome is~\cite{Selay2025WebTesting}. Metamorphic testing removes the last step by asking for a relation rather than an expected value~\cite{Chen2018Survey,Segura2016MTSurvey}, but until now the relation, the path, and the observables had to be written by a person who knew the site~\cite{Segura2019MRP,Chaleshtari2023Metamorphic}. AMoRE produces all three from a URL. In the study a compiled relation took a median of ten minutes of unattended exploration and about five cents of model usage, needed no page object, adapter, or expected output, and replayed in under half a minute with no model call. The tester's job now shifts from authoring tests to reviewing the drafts, and the review is informed by screenshots, traces, and observation values rather than by reading generated code.

The second effect is on oracle quality. LLM judgement as a regression oracle is contested because it is stochastic and struggles to separate real failures from harmless presentation changes~\cite{Dobslaw2026Challenges,Kaynak2025LLMShot}. AMoRE confines the model to proposing and lets deterministic code decide. The strict verdict is conservative and auditable. In the evaluation it exposed lost actions, divergent autocomplete outcomes, and logically contradictory observables that a permissive oracle would have accepted, and every verdict links back to the evidence that produced it. Because approved relations replay without inference, repeated runs measure the stability of the test and of the application rather than the variance of the model, which is what a regression suite needs.

AMoRE also opens research questions that were hard to study without an end-to-end pipeline. Since the model is a configuration value, different LLMs can be compared on identical prompts, inventories, and verdict code. Ablations can isolate the contribution of visual input, SoM grounding, fresh inventories after each probe, smoke replay, and family profiles. The quality categories observed in the study, namely strong, weak, brittle, misaligned, and non-evaluable relations, define gates that future work can automate and measure. Seeded faults in controlled applications would let the strict verdict be assessed as a fault detector rather than as a consistency check. The reviewer interface allows studies of how much effort the human gate costs and how often reviewers catch misaligned relations. The scripts, manifests, and per-relation review data of this study are published in the repository~\cite{AMoRERepo} as a starting point for such studies.

\section{Conclusions}
\label{sec:conclusions}

AMoRE demonstrates that an LLM-based agent can discover executable metamorphic relations for previously unseen web interfaces and compile them into Playwright tests that can be replayed without further model intervention. Across five commercial websites, 63 of 100 attempts compiled successfully and 48 produced evaluable source--follow-up pairs, at about USD 0.05 in model usage and a median of ten minutes of exploration per compiled relation.

The study also draws the boundary. Compilation does not guarantee a valid relation or informative observables, replay stability depends on how dynamic the page is, and a strict failure is a triage signal rather than proof of a defect. Human review at the promotion gate therefore remains part of the design.

Future work goes in three directions. Family-specific semantic checks and observable normalisation should reject invalid transformations, empty values, and wrongly directed comparisons before compilation, and a repository of reviewed relations could feed retrieval (RAG) during planning. Broader studies should cover more sites, authenticated flows, longer replay periods, several models, and a baseline of manually written tests, and should seed faults so that the verdict can be measured as an oracle. On the operational side, managed browser environments and a secure human takeover step would extend the tool to sites behind anti-bot gates.

\section*{Acknowledgements}

This work was carried out as part of the Master's thesis of the first author in the Master's programme in Software Engineering at the University of Murcia.

\bibliographystyle{elsarticle-num}
\bibliography{references}

@article{Weyuker1982,
  author  = {Weyuker, Elaine J.},
  title   = {On Testing Non-Testable Programs},
  journal = {The Computer Journal},
  volume  = {25},
  number  = {4},
  pages   = {465--470},
  year    = {1982},
  doi     = {10.1093/comjnl/25.4.465}
}

@article{Barr2015Oracle,
  author  = {Barr, Earl T. and Harman, Mark and McMinn, Phil and Shahbaz, Muzammil and Yoo, Shin},
  title   = {The Oracle Problem in Software Testing: A Survey},
  journal = {IEEE Transactions on Software Engineering},
  volume  = {41},
  number  = {5},
  pages   = {507--525},
  year    = {2015},
  doi     = {10.1109/TSE.2014.2372785}
}

@techreport{Chen1998,
  author      = {Chen, Tsong Yueh and Cheung, S. C. and Yiu, S. M.},
  title       = {Metamorphic Testing: A New Approach for Generating Next Test Cases},
  institution = {The Hong Kong University of Science and Technology},
  number      = {HKUST-CS98-01},
  year        = {1998}
}

@article{Chen2018Survey,
  author  = {Chen, Tsong Yueh and Kuo, Fei-Ching and Liu, Huai and Poon, Pak-Lok and Towey, Dave and Zhou, Zhi Quan and Zhang, Jian},
  title   = {Metamorphic Testing: A Review of Challenges and Opportunities},
  journal = {ACM Computing Surveys},
  volume  = {51},
  number  = {1},
  pages   = {4:1--4:27},
  year    = {2018},
  doi     = {10.1145/3143561}
}

@article{Segura2016MTSurvey,
  author  = {Segura, Sergio and Fraser, Gordon and Sanchez, Ana B. and Ruiz-Cort{\'e}s, Antonio},
  title   = {A Survey on Metamorphic Testing},
  journal = {IEEE Transactions on Software Engineering},
  volume  = {42},
  number  = {9},
  pages   = {805--824},
  year    = {2016},
  doi     = {10.1109/TSE.2016.2532875}
}

@inproceedings{Segura2019MRP,
  author    = {Segura, Sergio and Dur{\'a}n, Amador and Troya, Javier and Ruiz-Cort{\'e}s, Antonio},
  title     = {Metamorphic Relation Patterns for Query-Based Systems},
  booktitle = {Proceedings of the 4th International Workshop on Metamorphic Testing (MET@ICSE)},
  pages     = {24--31},
  year      = {2019},
  doi       = {10.1109/MET.2019.00012}
}

@article{Segura2018REST,
  author  = {Segura, Sergio and Parejo, Jos{\'e} A. and Troya, Javier and Ruiz-Cort{\'e}s, Antonio},
  title   = {Metamorphic Testing of {RESTful} Web {APIs}},
  journal = {IEEE Transactions on Software Engineering},
  volume  = {44},
  number  = {11},
  pages   = {1083--1099},
  year    = {2018},
  doi     = {10.1109/TSE.2017.2764464}
}

@article{Zhou2016SearchEngines,
  author  = {Zhou, Zhi Quan and Xiang, Sun and Chen, Tsong Yueh},
  title   = {Metamorphic Testing for Software Quality Assessment: A Study of Search Engines},
  journal = {IEEE Transactions on Software Engineering},
  volume  = {42},
  number  = {3},
  pages   = {264--284},
  year    = {2016},
  doi     = {10.1109/TSE.2015.2478001}
}

@article{Chaleshtari2023Metamorphic,
  author  = {Chaleshtari, Nazanin Bayati and Pastore, Fabrizio and Goknil, Arda and Briand, Lionel C.},
  title   = {Metamorphic Testing for Web System Security},
  journal = {IEEE Transactions on Software Engineering},
  volume  = {49},
  number  = {6},
  pages   = {3430--3471},
  year    = {2023},
  doi     = {10.1109/TSE.2023.3256322}
}

@inproceedings{Zhang2023ChatGPTMR,
  author    = {Zhang, Yi and Towey, Dave and Pike, Matthew},
  title     = {Automated Metamorphic-Relation Generation with {ChatGPT}: An Experience Report},
  booktitle = {2023 IEEE 47th Annual Computers, Software, and Applications Conference (COMPSAC)},
  pages     = {1780--1785},
  publisher = {IEEE},
  year      = {2023}
}

@misc{Shin2024Towards,
  author        = {Shin, Seung Yeob and Pastore, Fabrizio and Bianculli, Domenico and Baicoianu, Alexandra},
  title         = {Towards Generating Executable Metamorphic Relations Using Large Language Models},
  year          = {2024},
  eprint        = {2401.17019},
  archivePrefix = {arXiv},
  primaryClass  = {cs.SE}
}

@inproceedings{Tsigkanos2023Large,
  author    = {Tsigkanos, Christos and Rani, Pooja and M{\"u}ller, Sebastian and Kehrer, Timo},
  title     = {Large Language Models: The Next Frontier for Variable Discovery within Metamorphic Testing?},
  booktitle = {2023 IEEE International Conference on Software Analysis, Evolution and Reengineering (SANER)},
  pages     = {678--682},
  year      = {2023},
  doi       = {10.1109/SANER56733.2023.00070}
}

@inproceedings{Hazott2025LLM,
  author    = {Hazott, Christoph and Gro{\ss}e, Daniel},
  title     = {{LLM}-assisted Metamorphic Testing of Embedded Graphics Libraries},
  booktitle = {2025 Forum on Specification and Design Languages (FDL)},
  pages     = {1--10},
  year      = {2025},
  doi       = {10.1109/FDL68117.2025.11165405}
}

@misc{Khan2026Multi,
  author        = {Khan, Shehroz and Mughees, Abdullah and Sudheerbabu, Gaadha and Ahmad, Tanwir and Truscan, Dragos},
  title         = {Multi-Agent {LLM}-based Metamorphic Testing for {REST} {APIs}},
  year          = {2026},
  eprint        = {2605.28321},
  archivePrefix = {arXiv},
  primaryClass  = {cs.SE}
}

@article{Xu2024MR,
  author  = {Xu, Congying and Terragni, Valerio and Zhu, Hengcheng and Wu, Jiarong and Cheung, Shing-Chi},
  title   = {{MR-Scout}: Automated Synthesis of Metamorphic Relations from Existing Test Cases},
  journal = {ACM Transactions on Software Engineering and Methodology},
  volume  = {33},
  number  = {6},
  pages   = {150:1--150:28},
  year    = {2024},
  doi     = {10.1145/3656340}
}

@misc{Gao2025Assure,
  author        = {Gao, Xuanqi and Zhai, Juan and Ma, Shiqing and Xie, Siyi and Shen, Chao},
  title         = {Assure: Metamorphic Testing for {AI}-powered Browser Extensions},
  year          = {2025},
  eprint        = {2507.05307},
  archivePrefix = {arXiv},
  primaryClass  = {cs.SE}
}

@inproceedings{Liu2024GPTDroid,
  author    = {Liu, Zhe and Chen, Chunyang and Wang, Junjie and Chen, Mengzhuo and Wu, Boyu and Che, Xing and Wang, Dandan and Wang, Qing},
  title     = {Make {LLM} a Testing Expert: Bringing Human-like Interaction to Mobile {GUI} Testing via Functionality-aware Decisions},
  booktitle = {Proceedings of the 46th International Conference on Software Engineering (ICSE)},
  pages     = {100:1--100:13},
  year      = {2024},
  doi       = {10.1145/3597503.3639180}
}

@article{Liu2025Seeing,
  author  = {Liu, Zhe and Li, Cheng and Chen, Chunyang and Wang, Junjie and Chen, Mengzhuo and Wu, Boyu and Wang, Yawen and Hu, Jun and Wang, Qing},
  title   = {Seeing is Believing: Vision-Driven Non-Crash Functional Bug Detection for Mobile Apps},
  journal = {IEEE Transactions on Software Engineering},
  volume  = {51},
  number  = {12},
  pages   = {3452--3466},
  year    = {2025},
  doi     = {10.1109/TSE.2025.3614469}
}

@article{Wang2024Software,
  author  = {Wang, Junjie and Huang, Yuchao and Chen, Chunyang and Liu, Zhe and Wang, Song and Wang, Qing},
  title   = {Software Testing With Large Language Models: Survey, Landscape, and Vision},
  journal = {IEEE Transactions on Software Engineering},
  volume  = {50},
  number  = {4},
  pages   = {911--936},
  year    = {2024},
  doi     = {10.1109/TSE.2024.3368208}
}

@article{Dobslaw2026Challenges,
  author  = {Dobslaw, Felix and Feldt, Robert and Yoon, Juyeon and Yoo, Shin},
  title   = {Challenges in Testing Large Language Model Based Software: A Faceted Taxonomy},
  journal = {ACM Transactions on Software Engineering and Methodology},
  year    = {2026},
  note    = {Just Accepted},
  doi     = {10.1145/3806396}
}

@phdthesis{Selay2025WebTesting,
  author = {Selay, Elmin},
  title  = {On Automated Testing of Web Applications},
  school = {University of Wollongong},
  year   = {2025},
  doi    = {10.71747/uow-r3gk326m.31094914.v1}
}

@inproceedings{Kaynak2025LLMShot,
  author    = {Kaynak, Erg{\"u}n Batuhan and Lami, Mayasah and Moslemi, Sahand and Koyuncu, Anil},
  title     = {{LLMShot}: Reducing Snapshot Testing Maintenance via {LLMs}},
  booktitle = {2025 IEEE International Conference on Software Maintenance and Evolution (ICSME)},
  pages     = {827--832},
  year      = {2025},
  doi       = {10.1109/ICSME64153.2025.00087}
}

@misc{yang2023som,
  author        = {Yang, Jianwei and Zhang, Hao and Li, Feng and Zou, Xueyan and Li, Chunyuan and Gao, Jianfeng},
  title         = {Set-of-Mark Prompting Unleashes Extraordinary Visual Grounding in {GPT-4V}},
  year          = {2023},
  eprint        = {2310.11441},
  archivePrefix = {arXiv},
  primaryClass  = {cs.CV}
}

@misc{PlaywrightDocs,
  author = {{Microsoft}},
  title  = {Playwright Documentation: Fast and Reliable End-to-End Testing for Modern Web Apps},
  year   = {2026},
  url    = {https://playwright.dev/docs/intro},
  note   = {Accessed 14 September 2026}
}

@misc{LangGraphDocs,
  author = {{LangChain Inc.}},
  title  = {{LangGraph} Documentation},
  year   = {2026},
  url    = {https://langchain-ai.github.io/langgraphjs/},
  note   = {Accessed 14 September 2026}
}

@misc{AMoRERepo,
  author = {Coman, Andrei},
  title  = {{AMoRE}: An agentic framework for end-to-end metamorphic testing of web applications},
  year   = {2026},
  url    = {https://github.com/a-coman/AMoRE},
  note   = {Source code, validation scripts, and reproducibility data. Accessed 14 September 2026}
}

@book{brown2026c4model,
  title={The C4 Model: Visualizing Software Architecture},
  author={Brown, Simon},
  year={2026},
  publisher={O'Reilly Media},
  isbn={9798341660120}
}

@article{Baltes2026Guidelines,
  author  = {Baltes, Sebastian and Angermeir, Florian and Arora, Chetan and Mu{\~n}oz Bar{\'o}n, Marvin and Chen, Chunyang and B{\"o}hme, Lukas and Calefato, Fabio and Ernst, Neil and Falessi, Davide and Fitzgerald, Brian and Fucci, Davide and He, Junda and Treude, Christoph and Kalinowski, Marcos and Lambiase, Stefano and Russo, Daniel and Lungu, Mircea and {Martinez Montes}, Cristina and Prechelt, Lutz and Ralph, Paul and {van Tonder}, Rijnard and Wagner, Stefan},
  title   = {Guidelines for Empirical Studies in Software Engineering Involving Large Language Models},
  journal = {Empirical Software Engineering},
  volume  = {32},
  number  = {2},
  pages   = {64},
  year    = {2026},
  doi     = {10.1007/s10664-026-10922-3}
}

@misc{MiniMaxM3,
  author = {{MiniMax}},
  title  = {{MiniMax-M3} Model Card},
  year   = {2026},
  url    = {https://huggingface.co/MiniMaxAI/MiniMax-M3},
  note   = {Accessed 26 November 2026}
}

@article{Bons2023Scripted,
  author  = {Bons, Axel and Mar{\'i}n, Beatriz and Aho, Pekka and Vos, Tanja E. J.},
  title   = {Scripted and Scriptless {GUI} Testing for Web Applications: An Industrial Case},
  journal = {Information and Software Technology},
  volume  = {158},
  pages   = {107172},
  year    = {2023},
  doi     = {10.1016/j.infsof.2023.107172}
}

@article{SotoSanchez2022Dataset,
  author  = {Soto-S{\'a}nchez, {\'O}scar and Maes-Bermejo, Michel and Gallego, Micael and Gort{\'a}zar, Francisco},
  title   = {A Dataset of Regressions in Web Applications Detected by End-to-End Tests},
  journal = {Software Quality Journal},
  volume  = {30},
  number  = {2},
  pages   = {425--454},
  year    = {2022},
  doi     = {10.1007/s11219-021-09566-x}
}

\end{document}